# MAXI: all-sky observation from the International Space Station

Tatehiro Mihara[*a], Mutsumi Sugizaki[a], Masaru Matsuoka[a], Hiroshi Tomida[b], Shiro Ueno[b], Hitoshi Negoro[c], Atsumasa Yoshida[d], Hiroshi Tsunemi[e], Motoki Nakajima[f], Yoshihiro Ueda[g], Makoto Yamauchi[h]

[a] MAXI team, RIKEN, 2-1 Hirosawa, Wako, Saitama, 351-0198, Japan
[b] ISS Science Project Office, ISAS, JAXA, 2-1-1 Sengen, Tsukuba, Ibaraki, 305-8505, Japan
[c] Dept. of Physics, Nihon Univ., 1-8-14 Kanda-Surugadai, Chiyoda, Tokyo 101-8308, Japan
[d] Dept. of Phys. and Math., Aoyama Gakuin Univ., 5-10-1 Fuchinobe, Sagamihara, 252-5258, Japan
[e] Dept. of Earth and Space Sci., Osaka Univ., Machikaneyama, Toyonaka, Osaka 560-0043, Japan
[f] School of Dentistry at Matsudo, Nihon Univ., Sakaecho-nishi, Matsudo, Chiba 101-8308, Japan
[g] Dept. of Astronomy, Kyoto Univ., Kitashirakawa, Oiwake-cho, Sakyo-ku, Kyoto 606-8502, Japan
[h] Dept. of Applied Physics, Univ. of Miyazaki, Gakuen-Kibanadai-Nishi, Miyazaki 889-2192, Japan

**ABSTRACT**

Monitor of All-sky X-ray Image (MAXI) is mounted on the International Space Station (ISS). Since 2009 it has been scanning the whole sky in every 92 minutes with ISS rotation. Due to high particle background at high latitude regions the carbon anodes of three GSC cameras were broken. We limit the GSC operation to low-latitude region around equator. GSC is suffering a double high background from Gamma-ray altimeter of Soyuz spacecraft. MAXI issued the 37-month catalog with 500 sources above ~0.6 mCrab in 4-10 keV. MAXI issued 133 to Astronomers Telegram and 44 to Gamma-ray burst Coordinated Network so far. One GSC camera had a small gas leak by a micrometeorite. Since 2013 June, the 1.4 atm Xe pressure went down to 0.6 atm in 2014 May 23. By gradually reducing the high voltage we keep using the proportional counter. SSC with X-ray CCD has detected diffuse soft X-rays in the all-sky, such as Cygnus super bubble and north polar spur, as well as it found a fast soft X-ray nova MAXI J0158-744. Although we operate CCD with charge-injection, the energy resolution is degrading. In the 4.5 years of operation MAXI discovered 6 of 12 new black holes. The long-term behaviors of these sources can be classified into two types of the outbursts, 3 Fast Rise Exponential Decay (FRED) and 3 Fast Rise and Flat Top (FRFT). The cause of types is still unknown.

## 1. INTRODUCTION

MAXI is an X-ray all-sky monitor on the International Space Station (Matsuoka et al. 2009). It is equipped with Gas Slit Camera (GSC, Mihara et al. 2011) and Solid-state Slit Camera (SSC, Tomida et al. 2011). Since it was mounted to the Japanese experimental module in 2009, it has been scanning the whole sky in every 92 minutes with ISS rotation. The data are processed automatically and distributed through http://maxi.riken.jp homepage. In the following sections we describe GSC status, SSC status, and then MAXI discovery of black holes during 4.5 years of observation.

## 2. GSC

We operated whole 12 cameras in the standard voltage 1650V at first in 2009 August. Figure 1 shows the daily duty cycle of the operation in each camera. At first they were about 80 %. Two weeks later some cameras started heavy discharges. Location of the discharge on a carbon wire was concentrated at the edge of the wire frame. Finally one carbon wire was broken at the discharge position in GSC camera 6. The broken wire touches the surrounding ground wires and the bottom veto wires to short the high voltage. Since three carbon wires are connected, a half area (three out of six wires) became out of sensitivity. The rest half area does not have anti-coincidence by vetos, and the overall sensitivity of a camera became 5 times worse. Six days later one more camera (GSC camera 9) had the same damage. The discharge was caused by the high particle background at high earth-latitude regions. We limit the GSC operation to low-latitude region around equator within 40 degrees since 2009 September 23. The working time of GSC has become a half of the planned. Six month later, one more camera (GSC camera 3) had a damage of carbon wire at the center region, which was different position from the previous two. Then we decided to prevent even a small discharge. When a discharge was observed, we reduce the high voltage down, too. The high voltage of GSC camera 0 which had had a small discharge was set to 1550V. The voltages of other 3 cameras (GSC camera 7, 8, B) were also set down to 1550V to extend the life. Later on GSC camera 1 was set down. GSC camera 3 started the operation in 1550V but with a degraded sensitivity since 2010 June 21. Three cameras (GSC camera 1, 8, B) had a discharge even in 1550V and the operation has been tentatively stopped. Therefore, six cameras including one in a degraded sensitivity are working since 2012 April. Four cameras (GSC camera 1, 8, A, B) are standing by for 1500V operation. GSC camera 0 is standing by for a degraded sensitivity with 1550V. GSC camera 9 is off, since it cannot be operated continuously due to electric short.

---

[*] mihara@crab.riken.jp

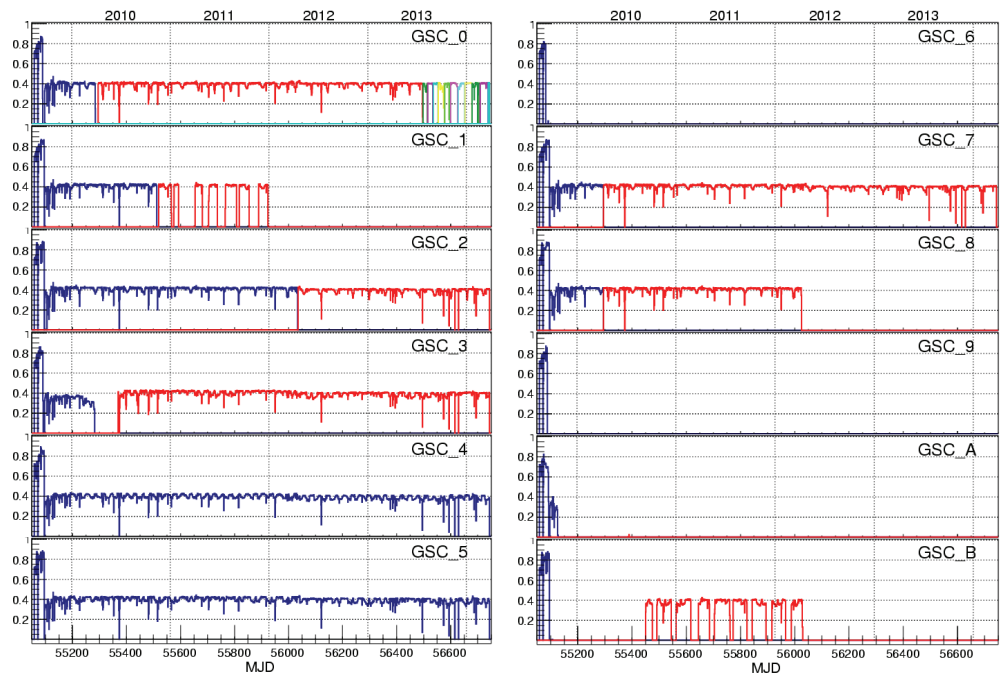

Figure 1. Daily observation duty cycle of MAXI GSC 12 cameras (GSC 0, GSC 1, ..., GSC B) for 4.5 years from 2009 August to 2014 April. Blue and red represent operations in high voltage of 1650V and 1550V, respectively. GSC 0 had a gas leak on 2013 June 15 and the high voltage is being reduced step by step.

Figure 2 is the background count rates in 4-10 keV band in the central region of the proportional counter. The anti-coincidence is functioning on board MAXI, and only the single (X-ray) events are sent to the ground. The rates sometimes go down to a half for tens of days. But the truth is that the background is kept twice high in most of the time. The high background spans are correlated to existence of a Soyuz spacecraft at the closest docking port. We found later that gamma-ray altimeter of Soyuz spacecraft has a strong isotope (~0.8 Ci of $^{137}$Cs and $^{60}$Co). It certainly doubles the background of MAXI/GSC. As you see in Figure 2, GSC 0 has least affected by the Soyuz, probably because it is mounted on the left side of the forward cameras which is the furthest corner from the Soyuz.

Figure 3 is the all-sky map with MAXI/GSC in 4.1 years. Although the Soyuz gamma-ray somewhat degrade the MAXI sensitivity, MAXI issued the 37-month catalog (Hiroi et al. 2012) which contains 500 sources above ~0.6 mCrab in 4-10 keV in high Galactic-latitude (|b| > 10 deg). In terms of rapid alerting, MAXI issued 133 to Astronomers Telegram and 44 to Gamma-ray burst Coordinated Network so far.

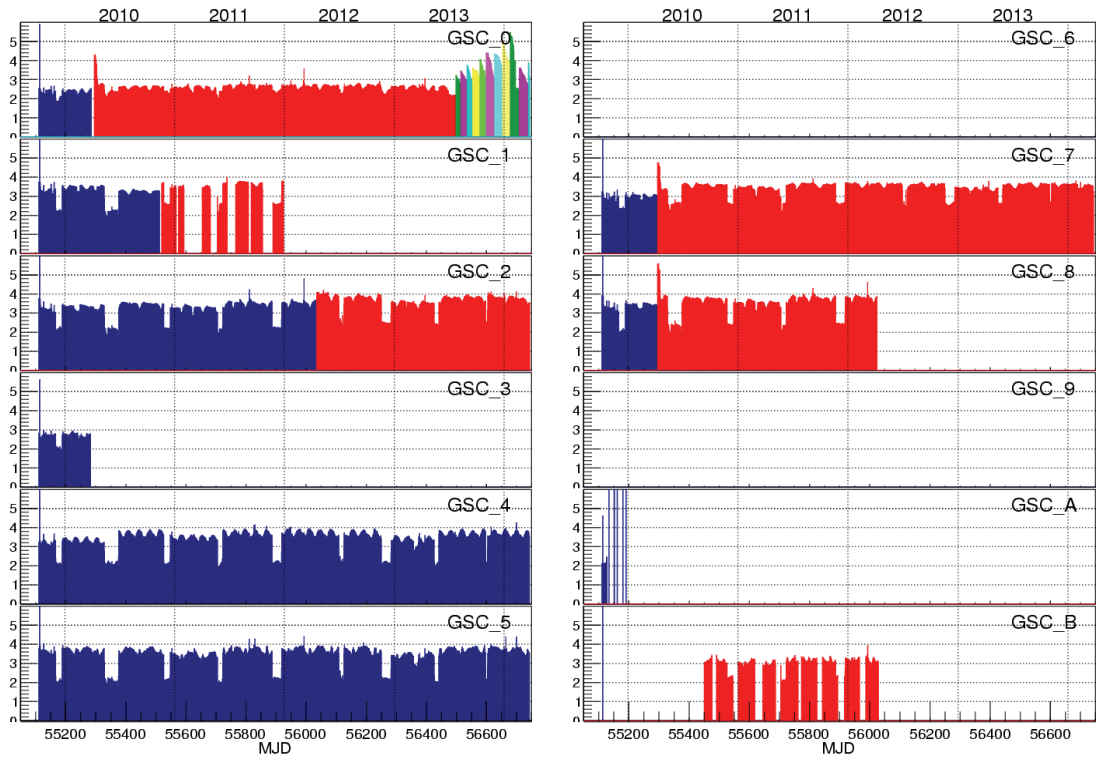

Figure 2. Background rates of 12 GSC counters for 4.5 years. Data represent the daily average per counter in 4-10 keV band. Blue and red represent operations in high voltage of 1650V and 1550V, respectively.

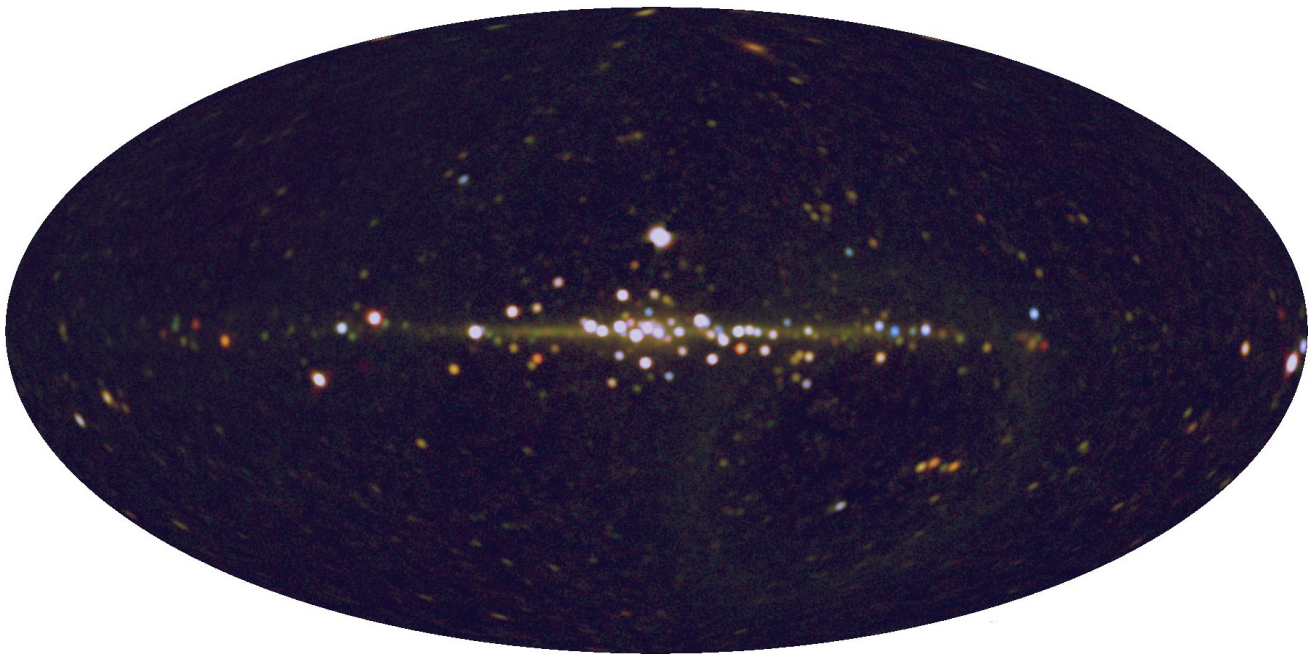

Figure 3. All-sky map with MAXI/GSC in 4.1 years. The color indications are Red: 2-4 keV, Green: 4-10 keV, and Blue: 10-20 keV. The X-ray binary pulsars appear in blue, supernova remnants in red. Yellows are low-mass X-ray binaries. More than 500 sources are detected.

On 2013 June 15, the gas gain of GSC camera 0 suddenly started to rise (Figure 4). A detailed discussion lead to a conclusion that the camera had a small gas leak probably caused by a hit of a small micrometeoride with about 50 micro meter in diameter. Using the gas gain equation in Knoll's handbook, we can estimate the pressure of the gas. The 1.4 atm Xe pressure went down to 0.6 atm in 2014 May 23. The diameter of the hole is estimated as ~4 micro meter on Be window of 100 micro meter thickness. We reduced the high voltage gradually and have kept the gas gain of the proportional counter. The sizes of the molecules of Xe (X-ray absorbing gas) and $CO_2$ (quenching gas) are almost the same, and the gas ratio would not change much. The stopping power of Xe in even 0.5 atm is enough to stop below 8 keV. The camera will keep functioning as long as we keep the gas gain.

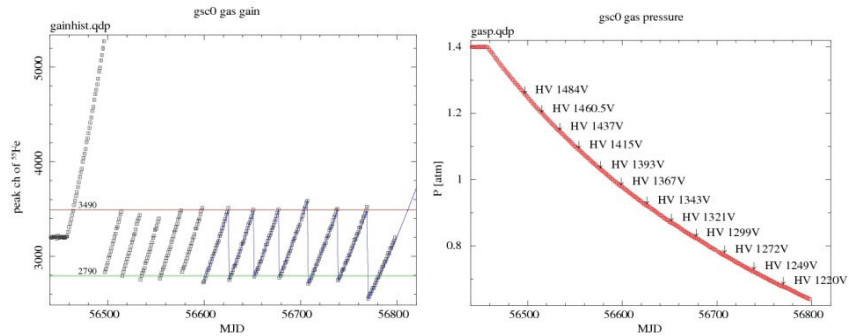

Figure 4. The peak pulse height of $^{55}$Fe calibration source of GSC camera 0. It suddenly started to rise on 2013 June 15. We noticed it one month later, and reduced the high voltage (from 1550V to 1484V) to bring it back to the original pulse height. We decided to keep the peak height between 2790 ch and 3490 ch (+- 12 %). We reduce the high voltage in every 3 weeks or a month. The right figure is estimated gas pressure by the pulse height change.

### 3. SSC

Figure 5 is the MAXI/SSC all-sky map in 2.5 years. Compared to GSC map, diffuse components extending to high-galactic latitude is apparent. Large SNR (Cygnus loop and Vela SNR) are also resolved. Figure 6 left is the spectrum of the supernova remnant Cas A. Emission lines of Si, S, Ar, and Fe are clearly detected. We have used the charge injection method from the beginning. Increase of the dark current, and degradation of charge transfer efficiency are observed. We corrected these effects and obtained an example of degradation of energy resolution in Figure 6 right. There was no degradation in the first year, but later on, the energy resolution is being degraded steadily.

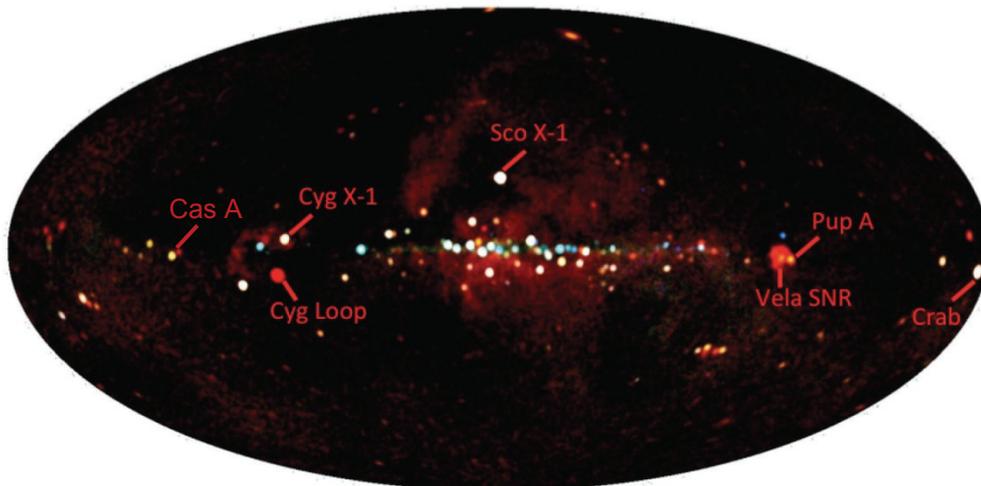

Figure 5. All-sky map with MAXI/SSC in 2.5 years. The color indications are Red: 0.7-1.7 keV, Green: 1.7-4.0 keV, and Blue: 4.0-7.0 keV.

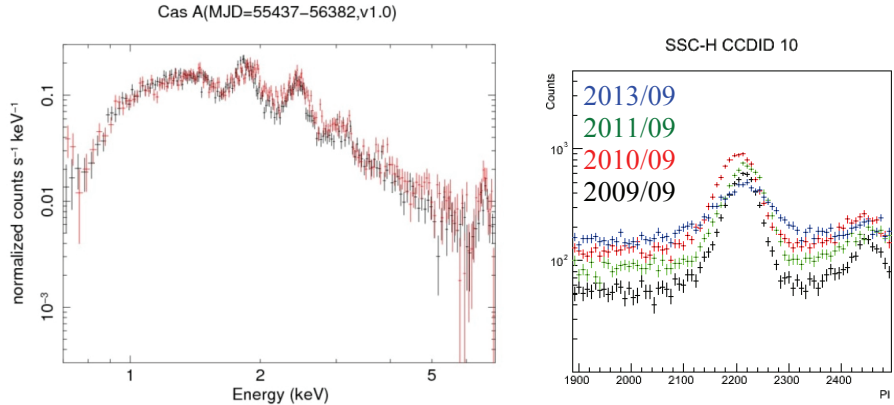

Figure 6. left: Spectrum of the supernova remnant Cas A with MAXI/SSC from 2010 August 29 to 2013 March 31. Emission lines of Si, S, Ar, and Fe are clearly detected. Right : Spectra of an instrumental Cu K alpha line (8.04 keV) of a CCD in MAXI/SSC. Degradation of the energy resolution in the four years is seen.

SSC with X-ray CCD has detected diffuse soft X-rays in the all-sky, such as Cygnus super bubble (Kimura et al. 2013) and north polar spur, as well as it found a fast soft X-ray nova MAXI J0158-744 (Morii et al. 2013).

## 4. MAXI DISCOVRY OF BLACK HOLES

Since the beginning of the MAXI mission (August 2009), 12 BHCs have been discovered. MAXI discovered 6 of them solely or independently. They are MAXI J1659-152, MAXI J1543-564, MAXI J1836-194, MAXI J1305-704, MAXI J1910-057/Swift J1910.2-0546, and MAXI J1828-249. The light curves from the discovery are shown in Figure 7. One of the most characteristic features in the MAXI BHCs is their faintness. No bright outbursts with 2-10 keV fluxes of more than 1 Crab was not observed, while they were often observed with Ginga and RXTE. Only the 2-4 keV flux of J1910-057 exceeded 1 Crab slightly. J1543-564, J1305-704 and J1828-249 showed soft state transitions, but their peak fluxes were only 100 mCrab, only 1/100 of those of Ginga X-ray novae (see Figure 3 in Tanaka & Shibazaki 1996). If the state transition is simply determined by an Eddington rate (e.g. Maccarone 2003), those sources must be 10 times further than the Ginga nova.

As shown in Figure 7 left and right, there seems to be two kinds of long-term variations after the maximum in the MAXI sources. One is a classical, "fast rise and exponential decay, FRED (Chen et al. 1997)" type of the outbursts: MAXI J1836-194, J1901-057, and J1828-249 (Figure 7 left). Such outbursts were commonly observed in the Ginga era, but not so common in the RXTE era (e.g. McClintock & Remillard 2006). The time profiles of J1836-194 and J1828-249 are very similar, but the spectral evolutions are different. J1828-249 showed the soft state transition rapidly, but J1836-194 failed. If the difference simply comes from the accretion rate, J1828-249 must be further than J1836-194 though both are at similar galactic latitudes ($b$ = -6.5 deg and $b$ = -5.4 deg, respectively). It should be also noted that the decay time constants of these three BHCs are about 25 days, though the decay constants are influenced by spectral changes and depend on the term to fit. Similar decay time constants are obtained in A 0620-00 ( 24 days), GS 2000+251 ( 30 days), and GS 1124-684 (30 days) (see Tanaka & Shibazaki 1996, references therein) and others (Chen et al. 1997). The cause of the exponential decay has not been fully understood. The constancy of the decay constants in BHCs with various masses implies that the decay constant hardly depends on the mass.

The other group of the outbursts is a "fast rise and flat-top, FRFT" type of the outbursts : MAXI J1659-152, J1543-564, and J1305-704 (Figure 7 right). These sources showed flux plateaus lasting 30-60 days. Note that such morphology strongly depends on the energy band (we observe different components in different energy bands), and it should be done in the same energy band. Also note that in both the types of the outbursts, the flux changed by a factor of about two near the peak, resulting in multiple peaks, which can be clearly seen when plotted in a linear scale.
The flux plateau implies a relatively constant accretion rate during it, which might result from Roche-lobe overflow rather than temporary gas accretion due to thermal instability of the accretion disk (Mineshige & Wheeler 1989). It is interesting to note that a long-term profile of J1659-152 stretching to the time-axis direction is similar to those of J1305-704 and J1543-564 (Fig. 7 right). This might simply reflect the size of the binary system. The shortest orbital period and an unobservable companion star in J1659-152 imply the system to be the most compact with a small size of the accretion disk and a small companion star, resulting in a small reservoir to provide accretion matter.

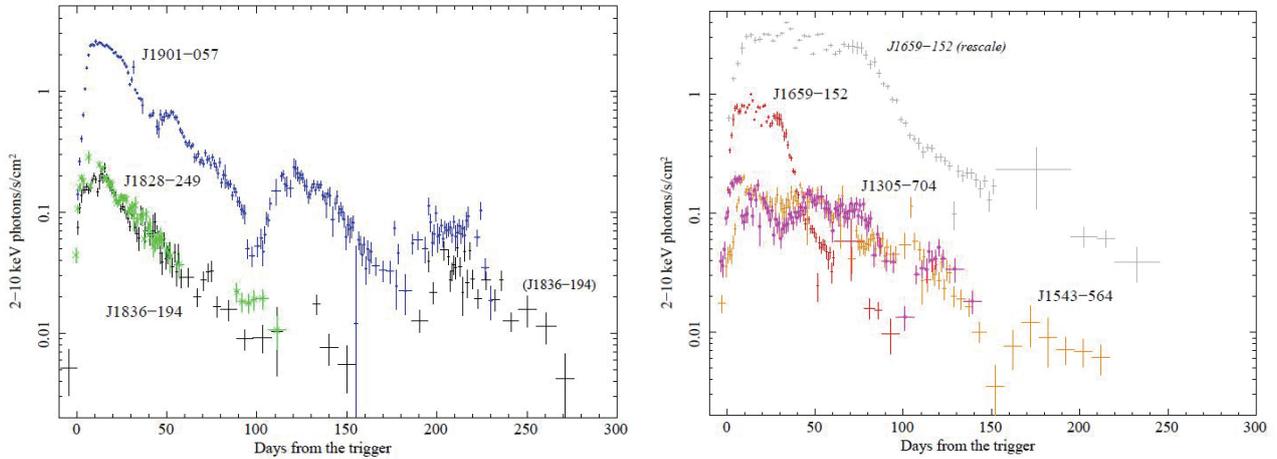

Figure 7. left: 2-10 keV light curves of 3 FRED type X-ray novae, J1836-194 (black) J1901-057 (blue and circles), and J1828-249 (green and crosses ). Right: 2-10 keV light curves of 3 FRFT type X-ray novae, J1659-152 (red) J1543-564 (orange), and J1305-704 (Magenta and circles). The rescaled light curve of J1659-152, stretching by 2.5 times to the horizontal direction and 4 times to the vertical one was also shown (light grey).